\documentstyle[12pt]{article}
\begin{document}
\begin{titlepage}
\rightline{APCTP-97-10}
\rightline{UM-TG-194}
\def\today{\ifcase\month\or
        January\or February\or March\or April\or May\or June\or
        July\or August\or September\or October\or November\or December\fi,
  \number\year}
\rightline{hep-th/9705106}
\vskip 1cm
\centerline{\Large \bf Geometry, D-Branes and  $N=1$ Duality}
\centerline{\Large \bf in 
Four Dimensions with Product Gauge Groups}
\vskip 2cm
\centerline{\sc Changhyun Ahn \footnote{chahn@apctp.kaist.ac.kr}}
\centerline {{\it Asia Pacific Center for Theoretical Physics (APCTP)}}
\centerline{{\it 207-43 Cheongryangri-dong, Dongdaemun-gu}}
\centerline {{\it Seoul 130-012, Korea}}
\centerline{and}
\centerline{ \sc Radu Tatar \footnote{tatar@phyvax.ir.miami.edu}}
\centerline{\it Dept. of Physics}
\centerline{ \it University of Miami}
\centerline{ \it Coral Gables, Florida 33146, USA}
\vskip 2cm
\centerline{\sc Abstract}
\vskip 0.2in
We study $N=1$ dualities in four dimensional supersymmetric gauge theories
by D 6-branes wrapping around 3-cycles of Calabi-Yau threefolds in type IIA
string theory. We consider the models involving $SU(N_{c1}) \times 
SU(N_{c2})$ product gauge group without or with adjoint matter in terms of
geometrical realization of the configuration of D 6-branes wrapped 3-cycles.
We also find simple geometric descriptions for the triple product gauge group
$SU(N_{c1}) \times SU(N_{c2}) \times SU(N_{c3})$ interpreted recently by
Brodie and Hanany in the context of D brane configurations together with
NS 5-branes. Their introduction of semi infinite D 4-branes appears
naturally by looking at the flavor group in the dual theory. 
We generalize to product of arbitrary  number of gauge groups.    
 
\vskip 0.6in
\leftline{May, 1997}
\end{titlepage}
\newpage
\section{Introduction and Geometrical Picture}
String theory interprets many nontrivial aspects of four dimensional
$N=1$ supersymmetric field theory by T-duality on the
local model for the compactification manifold.
There are two approaches to describe this local model. 

{ \it One} is to consider the local description as purely geometric
structure of compactification manifold together with D-branes wrapping
around cycles \cite{KV1,BJPSV,VZ}.
The compactification of F-theory on elliptic 
Calabi-Yau(CY) fourfolds from 12 dimensions leads to
$N=1$ supersymmetric field theories in four dimensions.
It has been studied in \cite{KV1},
for the case of pure $SU(N_c)$ Yang-Mills gauge theory,
that the gauge symmetry can be obtained in terms of the structure of
the D 7-brane worldvolume.
By adding D 3-branes and bringing them
near the complex 2-dimensional surface,   
the local string model gives rise to matter hypermultiplets in the 
fundamental representation with
pure $SU(N_c)$ Yang-Mills theory \cite{BJPSV}. 
Moreover, Seiberg's duality \cite{Seiberg} 
for the $N=1$
supersymmetric field theory can be mapped to a T-duality exchange of the
D 3-brane charge and D 7-brane charge. 
For the extension to $SO(N_c)$ and $Sp(N_c)$ 
gauge theories \cite{IS,IP} coupled
to matter, the local string models are type IIB orientifolds 
with D 7-branes on a curved orientifold
7-plane \cite{VZ}.

{\it The other}  is 
to interpret $N=1$ duality for $SU(N_c)$ gauge theory
as D-brane description together with
NS 5-branes in a flat geometry \cite{EGK} according 
to the approach of \cite{HW}. 
Extension of this to the case of $SO(N_c)$ and $Sp(N_c)$
gauge theories with flavors was presented in \cite{EJS} by
adding an orientifold 4-plane.
The generalization to the construction of product gauge 
groups $SU(N_{c1}) \times
SU(N_{c2})$ with matter fields is given in \cite{bh,BSTY} by 
suspending two sets of
D 4-branes between three NS 5-branes.
It was also considered in \cite{Tatar} a  
brane configuration  which gives rise to electric and magnetic theories for
$ SO(N_{c1}) \times Sp(N_{c2})$ 
product gauge group. Very recently, the dual aspects of $N=1$ 
supersymmetric $SO(10)$ gauge theory with arbitrary numbers of spinors
and vectors have been found in \cite{BCKS} in the context of field theory.

Following the idea of \cite{BJPSV,VZ},
Ooguri and Vafa have observed in \cite{ov} that
$N=1$ duality can be embedded into 
type IIA string theory with D 6-branes, partially
wrapped around three cycles of CY threefold, filling four
dimensional spacetime.
They discussed 
what happens to the wrapped cycles and studied
the relevant field theory results when a transition in  the moduli of CY 
threefolds occurs. Furthermore, they reinterpreted the configuration of
D-branes in the presence of NS 5-branes \cite{EGK} 
as purely classical geometrical
realization.

It is natural to ask how a number of known field theory dualities
which contain {\it additional} field contents
arise from two different approaches we have described so far. 
It was obtained in \cite{ca1}, according to the approach of \cite{ov}, that 
one can  generalize the work of \cite{ov} to various models,
consisting of one or two 2-index tensors 
and some fields in the defining
representation (fundamental representation for $SU(N_c)$ and $Sp(N_c)$,
vector representation for $SO(N_c)$), presented earlier by many 
authors \cite{Kutasov} and study its geometric realizations
by wrapping D 6-branes about 3-cycles of CY threefolds. 
On the other hand, by introducing a multiple of coincident NS 5-branes
and  NS' 5-branes, it was argued \cite{EGKRS} that $SU(N_c)$ gauge
theory with one or two adjoints superfields 
in addition to the fundamentals can be
obtained. Moreover, $SO(N_c)$ gauge group 
with an adjoint field and $N_f$ vectors
was described in terms of a multiple of 
coincident NS 5-branes and a single NS'
5-brane with orientifold 4-plane ( similarly $Sp(N_c)$ 
with a traceless antisymmetric
tensor and $N_f$ flavors by adding orientifold 6-plane).
See also the relevant papers \cite{AH} 
analyzing the brane configurations
associated with field theories in various dimensions.
Recently \cite{ca2} we applied the approach of \cite{ov} to the various models
presented by Brodie and Strassler \cite{BS}, consisting of 
$D_{k+2}$ superpotential with $SU, SO$ and 
$Sp$ gauge groups along the line of
\cite{ca1}.
We discussed a large number of representations 
for a field $Y$, but with a field $X$
always in the adjoint (symmetric) [antisymmetric] 
representation for $SU (SO) [Sp]$
gauge groups where the superpotential $W= \mbox{Tr} X^{k+1}+
\mbox{Tr} XY^2$. 

In this paper, we go one step further. We find geometric descriptions for 
the multiple product gauge groups for $SU(N_c)$ without or with adjoint
matter. We compare our results with those obtained by Brodie and Hanany 
\cite{bh}
who used a different approach and reinterpret their semi infinite 
D 4-branes as the condition for flavor gauge group in the dual picture.   

We will  review the main geometrical 
setup of \cite{ov,ca1} in the remaining part of
this section.
Let us start with the compactification of type IIB string theory
on the CY threefold leading to $N=2$ supersymmetric field theories in 4 
dimensions. Suppose we have various D 3-branes wrapping around
a set of three cycles of CY threefold. It is 
known \cite{CDFV} that whenever
the integration of the holomorphic 3-form on the CY threefold 
around three cycles  
takes the form of parallel vectors in the complex plane, 
such a D 3-brane configuration
allows us to have a BPS state.
Then after we T-dualize the 3-spatial directions of three torus
$T^3$ we obtain type IIA string theory with D 6-branes, partially
wrapping around three cycles of CY threefold and
filling 4 dimensional spacetime. 
We end up with 
$N=1$ supersymmetric field theories in 4 dimensions. 

The local model of CY threefold can be described by \cite{BSV,ov}
five complex coordinates $x, y, x', y'$ and $z$ satisfying the following
equations:
$$
x^2+y^2=\prod_i (z-a_i), \;\;\;\; x'^2+y'^2=\prod_j (z-b_j)
$$ 
where each of $C^*$'s is embedded in $(x, y)$-space and $(x', y')$-space
respectively over a generic point $z$. 
This describes a family of a product of
two copies of one-sheeted hyperboloids in $(x, y)$-space 
and $(x', y')$-space respectively parameterized
by the $z$-coordinates. 
For a fixed $z$ away from $a_i$ and $b_j$ there
exist nontrivial $S^1$'s in each of $C^*$'s corresponding to the waist
of the hyperboloids. Notice that
when $z=a_i$ or $z=b_j$ the corresponding circles 
vanish as the waists shrink.
Then we regard 3 cycles as the product of $S^1 \times S^1$ cycles over
each point on the $z$-plane, with the segments in the $z$-plane ending
on $a_i$ or $b_j$. When we go between two $a_i$'s ($b_j$'s) 
without passing through
$b_j$ ($a_i$) the 3 cycles sweep out $S^2 \times S^1$.
On the other hand, when we go between $a_i$ and $b_j$ the 3 cycle becomes
$S^3$. We will denote the 3-cycle lying over between $a_i$ and $b_j$
by $[a_i, b_j]$ and also denote other cycles in a similar fashion.
In next section we will describe our 
main results by exploiting this geometrical setup.
We will start with by writing down 
the configurations of ordered points in the 
real axis of $z$-plane to various models
we are concerned with.

\section{ Two gauge groups: $SU(N_{c1}) \times SU(N_{c2})$}

Let us first briefly review the product gauge group 
$SU(N_{c1})\times SU(N_{c2})$
in the brane configuration picture. As discussed in \cite{bh}, in this case it
does not suffice to have NS 5 branes oriented at 0 degrees (i.e. in $(x^{4}, 
x^{5})$ plane) and 90 degrees (i.e. in $(x^{8}, x^{9})$ plane). We cannot have
parallel NS 5-branes for a theory which has N=1 supersymmetry, 
so for more than two NS 5-branes, they need to appear at
different angles in $(x^{4}, x^{5}, x^{8}, x^{9})$. To be more specific we 
repeat here an argument of \cite{bar}. One defines the complex planes 
$z_{1}= x^{4}+i x^{8}$ and $z_{2}= x^{5}+ i x^{9}$. 
In the case of only two NS 5-branes
, it is enough to have branes in $(x^{4}, x^{5})$ direction (which is the one
given by Im $z_{1} =$ Im $z_{2} = 0$ and in $(x^{8}, x^{9})$ 
direction (which is 
the one given by Re $z_{1} =$ Re $z_{2} = 0$). If one has more than 
two NS 5-branes,
a configuration which preserves exactly $N=1$ supersymmetry is the one in which
all the NS 5-branes are rotated at different angles in $(x^{4}, x^{5}, x^{8},
x^{9})$ direction. This can be understood by applying a rotation 
$z_{1}\rightarrow e^{i\theta_{i}}z_{1}$ and $z_{2}\rightarrow e^{-i\theta_{i}}
z_{2}$ to the $i$-th NS 5-brane.

Using this argument, it was considered in \cite{bh} three 
NS 5-branes A, B and C from left
to right with $N_{c1}$ D 4-branes 
between A and B and $N_{c2}$ D 4-branes between
B and C.  
A is rotated at an angle $\theta_{1}$,
B is in $(x^{4}, x^{5})$ direction and C is rotated at an angle $\theta_{3}$.
They also inserted $N_{f1}$ D 6-branes parallel with A and $N_{f2}$ 
D 6-branes 
parallel with C. This is the electric theory. The magnetic theory was obtained
by moving B and C to the right of A and in the final picture there were
$\widetilde{N}_{c1} (= 2 N_{f2}+N_{f1}-N_{c2})$ D 4-branes between C and B and
$\widetilde{N}_{c2} (= 2 N_{f1}+N_{f2}-N_{c1})$ D 4-branes between B and A.

Now we go to the geometric approach and we study how the duality is
obtained. 
In the geometric approach there are two possibilities to view the theory, one
is to take the original geometric picture 
in terms of singularities of CY 3-folds and the other one is obtained by a
T-duality applied to the original picture and is almost
identical with the description of the brane configuration approach. 
The last one differs from the description of the
brane configuration approach in the sense that
the role of the $N_{f}$ D 6-branes
is played by a NS 5-brane. To be more precise, in brane configuration,
we consider two NS 5-branes A and B with $N_{c}$ D 4-branes 
connecting them and 
$N_{f}$ D 6-branes which intersect the D 4-branes. 
On the other hand in the geometric approach, after a T-duality,
we have three NS 5-branes A, B and C, $N_{c}$ D4 branes
between A and B and $N_{f}$ D4-branes between B and C where C 
has to be parallel with A.

Let us introduce now the electric theory in the geometric approach. 
We study supersymmetric Yang-Mills theory with gauge group 
$SU(N_{c1})\times SU(N_{c2})$ , with $N_{f1}(N_{f2})$ flavors 
in the fundamental 
representation of the
first(second) gauge group. 
Besides we have the fields $X$ and $Y$ in the $\bf {(N_{c1}, 
\overline{N}_{c2})}$
and $\bf {(\overline{N}_{c1}, N_{c2})}$ 
representations of the product gauge groups.
The theory has a superpotential W = Tr $(XY)^{2}$
for unique NS 5-branes localised at different singularity points and
W = Tr $(XY)^{k+1}$ for $k$ NS 5-branes localised at every singularity point.

{\bf 2.1) The theory with superpotential W = Tr $(XY)^{2}$}

We start by wrapping D 6-branes around 
3-cycles of a Calabi-Yau threefold in type IIA string theory where
the 3-cycle is either $S^{2}\times S^{1}$ or $S^{3}$. We may 
regard $(x^{6}, x^{7})$ as real and imaginary
parts of $z$ coordinate. 
The geometry is viewed as a $C^{*}\times C^{*}$ bundle over the 
$z$-plane with $(x, y)$ coordinates in the 
first $C^{*}$ and $(x', y')$ coordinates in
the second $C^{*}$.
On the real axis of $z$-plane, we consider 5 singular 
points denoted by $( a_{1}, c_{2}, b, 
a_{2}, c_{1} )$ from left to right.  We shall later see (after T-duality)
why we make this choice. Suppose that the first $C^{*}$ degenerates at
$a_{1}, a_{2}, c_{1}$ and $c_{2}$ and the second $C^{*}$ degenerates at $b$.
In the fibre above $b$, 
the second $C^{*}$ is in (45) direction while the fibre above 
$a_{1}$ and $a_{2}$ is rotated at an angle $\theta_{1}$ in (4589) directions
and the fibre above $c_{1}$ and $c_{2}$ is rotated at an angle 
$\theta_{2}$ in (4589) directions. 
The cycles $[a_{i}, b],[c_{i}, b]$ 
are $S^{3}$ cycles and the cycles $[a_{1}, a_{2}], [c_{1}, c_{2}]$ and 
$[a_{i}, c_{j}]$ are $S^{2}\times S^{1}$ cycles. We wrap $N_{c1}$ D 6-branes
on the cycle $[a_{1}, b]$, $N_{c2}$ D 6-branes on the cycle $[b, c_{1}]$,
$N_{f1}$ D 6-branes on the cycle $[b, a_{2}]$ and $N_{f2}$ D 6-branes on
the cycle $[c_{2}, b]$. The field theory we end up with is given by an
$N=1$ gauge theory with gauge group $SU(N_{c1})\times SU(N_{c2})$ with 
flavor groups $SU(N_{f1})$ and $SU(N_{f2})$ which now are gauge groups. 
In order
to make them global we need to push $a_{2}$ and $c_{2}$ to infinity.
To do that we have to interchange the positions of $a_{2}$ and $c_{1}$ on 
one side and the positions of
$a_{1}$ and $c_{2}$ on the other side. 

When we make this change something interesting happens. To see that
let us perform
a T-duality in $(x^{4}, x^{5}, x^{8}, x^{9})$ directions. 
We now have five NS 5-branes, one at each of the previous five
points. The ones at $a_{1}$ and $a_{2}$(those at $c_1$ and $c_2$) are 
parallel and are rotated at an angle
$\theta_{1}(\theta_2)$ in (4589) directions. The NS 5-branes
at $a_{2}$ and $c_{2}$ play the role of the $N_{f1}$ and $N_{f2}$ D 6-branes
respectively
in the brane configuration picture where by
moving $N_{f1}$ D 6-brane from right to left with respect to a NS 5-brane, 
$N_{f1}$ D 4-brane 
will appear to the left of the $N_{f1}$ D6-brane. 
Our claim here is that when we move
the NS 5-branes sitting at $c_{2}$ from right to left with respect to the one
sitting at $a_{1}$, the same amount of D 4-branes will appear! This is a new
phenomenon compared with the previous considered cases and is the main idea
of our present work. It has to occur in order to match the results obtained
in field theory and in brane configuration. 

Going back (by T-duality) to the original picture, this means that when we move
$c_{2}$ to the left of $a_{1}$ then 
$N_{f2}$ supplementary D 6-branes are wrapped
on $[c_{2}, a_{1}]$. The same argument tells us that 
when we move $a_{2}$ to the
left of $c_{1}$, then $N_{f1}$ supplementary D 6-branes are wrapped on
$[c_{1}, a_{2}]$. We thus have the following configuration 
( from left to right):
$2N_{f2}$ D 6-branes are wrapped between
$c_{2}$ and $a_{1}$, $(N_{c1} + N_{f2})$ D 6-branes are wrapped 
between $a_{1}$ and $b$, $(N_{f1} + N_{c2})$ D 6-branes are wrapped between 
$b$ and $c_{1}$ and
$2 N_{f1}$ D 6-branes are wrapped between $c_{1}$ and $a_{2}$. 

Now we want to move to other point in the moduli of the CY threefold and end up
with a configuration where again the degeneration points are 
along the real axis
in the $z$-plane, but the order is changed to 
$(c_{2}, c_{1}, b, a_{1}, a_{2})$.
The transition is  similar to the one made for simple groups, in the sense that
 the point $b$ changes positions with both $a_{1}$ and $c_{1}$. This is
the reason for taking
the original picture the way we took it. First we push $c_{1}$ up along the
imaginary direction. Then $(N_{f1} + N_{c2})$ D 6-branes are wrapped on
$[b, a_{2}]$ and $(N_{f1} - N_{c2})$ D 6-branes are wrapped on 
$[c_{1}, a_{2}]$. We continue
to move $c_{1}$ along the negative real axis and insert it between $c_{2}$ and
$a_{1}$. At this moment the $(N_{f1} - N_{c2})$ D 6-branes 
which were wrapped on
$[c_{1}, a_{2}]$ decompose to $(N_{f1}-N_{c2})$ D 6-branes wrapped on
$[c_{1}, a_{1}]$,
$(N_{f1}-N_{c2})$ D 6-branes wrapped on $[a_{1}, b]$ and 
$(N_{f1} - N_{c2})$ D 6-branes wrapped on
$[b, a_{2}]$. So, after the first transition, there are $2N_{f2}$ D 6-branes 
wrapped on $[c_{2},c_{1}]$, $(2N_{f2}+N_{f1}-N_{c2})$ D 6-branes on 
$[c_{1}, a_{1}]$,
$(N_{c1} + N_{f1} + N_{f2} -N_{c2})$ D 6-branes wrapped on
 $[a_{1}, b]$ and $2N_{f1}$ along $[b, a_{2}]$. 
Similarly we push $b$ between $c_{1}$ and $a_{1}$. When we raise it along the 
imaginary axis there are $(N_{c1} + N_{f1} + N_{f2} -N_{c2}) $ 
D 6-branes wrapped 
on $[a_{1}, a_{2}]$ and $(N_{f1} - N_{f2} -N_{c1} + N_{c2})$ 
D 6-branes wrapped on
$[b,a_{2}]$. By moving the point 
$b$ along the negative real axis and inserting it 
between $c_{1}$ and $a_{1}$, we obtain the final configuration of points
ordered as $(c_{2}, c_{1}, b, a_{1}, a_{2})$ with $2N_{f2}$ D 6-branes 
wrapped on
$[c_{2}, c_{1}]$, $(2N_{f2} + N_{f1}-N_{c2})$ D 6-branes wrapped on
$[c_{1}, b]$, $(2N_{f1} + N_{f2} -N_{c1}) $ D 6-branes wrapped on 
$[b, a_{1}]$ and $2 N_{f1}$ D 6-branes wrapped on $[a_{1}, a_{2}]$.
This picture describes the magnetic theory with a gauge group
$SU(2N_{f2} + N_{f1} - N_{c2})\times SU(2N_{f1} + N_{f2} - N_{c1})$ with
$N_{f2}(N_{f1})$ flavors in the fundametal representation of 
the first(second) gauge group.
In order to obtain global flavor group, we move $a_{2}$ and $c_{2}$
to infinity. Besides we have singlets which correspond to the mesons of the
original electric theory and interact with the dual quarks through the
superpotential in the magnetic theory.
The magnetic description coincides with the one obtained in field theory
and in brane configuration.

{\bf 2.2) The theory with superpotential W = Tr $(XY)^{k+1}$}

In the brane configuration, the electric theory corresponds to
$k$ NS 5-branes with  the same orientation as A connected to a 
single B NS 5-brane by
$N_{c1}$ D4-branes from the left and $k$ NS 5-branes 
with the same orientation as
C connected to a single B NS 5-brane from the right by $N_{c2}$ D 4-branes.
 In the geometrical picture, after a T-duality, this would
mean that instead of having one NS 5-brane at $a_{1}$ we have
$k$ NS 5-branes and instead of having one NS 5-brane at $c_{1}$ we have
$k$ NS 5-branes. In order to handle the problem we consider instead
of $k$ NS 5-branes at one point, $k$ points with one NS 5-brane at each point.
This way of seeing things was introduced in \cite{ca1} for the group
$SU(N_{c})$ with adjoint matter.
The electric theory is a $SU(N_{c1})\times SU(N_{c2})$ gauge group with
$N_{f1}(N_{f2})$ flavors in the fundamental representation of the 
first(second) gauge group. 
Besides we have the same
fields $X$ and $Y$ as in the previous subsection with the
superpotential W = Tr $(XY)^{k+1}$.

For simplicity, let us consider first the case $k=2$. On the real axis we will
have from left to right: $(a_{11}, a_{12}, c_{2}, b, a_{2}, c_{11}, c_{12})$,
where instead of $a_{1}(c_1)$ we have $a_{11}, a_{12}(c_{11}, c_{12})$. 
Using the idea of \cite{ca1,ca2}, we wrap $N_{1}$
D 6-branes on $[a_{11}, b]$, $N_{2}$ D 6-branes on $[a_{12}, b]$, 
$N_{3}$ D 6-branes on $[b, c_{11}]$ and $N_{4}$
on $[c_{12}, b]$ 
such that $N_{1} + N_{2} = N_{c1}$ and $N_{3} + N_{4} = N_{c2}$. Besides
we wrap $N_{f1}$ D 6-branes on $[b, a_{2}]$ and $N_{f2}$ D 6-branes 
on $[b, c_{2}]$. As before, 
we move $c_{2}$ to the left of $a_{11}, a_{12}$ and $a_{2}$ to the right of
$c_{11}$ and $c_{12}$. 

To apply the observation made in the previous subsection, 
perform first a T-duality
in $(x^{4}, x^{5}, x^{8}, x^{9})$ directions. In the T-dual language,
to move $c_{2}$ to the left of $a_{11}, a_{12}$ means to move the 
NS 5-brane sitting at $c_{2}$ to the left with respect to the 
NS 5-branes sitting
at $a_{11}$ and $a_{12}$. When $c_{2}$
passes $a_{11}$, $N_{f2}$ supplementary D 6-branes 
are wrapped on $[c_{2}, a_{11}]$
and when it passes $a_{12}$ another $N_{f2}$ D 6-branes appear! 
Going back to the
original geometrical picture, we have
$3N_{f2}$ D 6-branes wrapped on $[c_{2}, a_{11}]$, $(2N_{f2} + N_{1})$ 
D 6-branes wrapped on
$[a_{11}, a_{12}]$ and $(N_{f2} + N_{c1})$ D 6-branes wrapped on 
$[a_{12}, b]$. In the limit
$a_{11}\rightarrow a_{12}$, the only difference between this configuration and
the one of the previous subsection is that we have 
$3N_{f2}$ D 6-branes wrapped on
$[c_{2}, a_{11} = a_{12} = a_{1}]$ instead of $2 N_{f2}$.
The same thing appears when we move $a_{2}$ to the right of $c_{11}, c_{12}$.
In the limit $c_{11}\rightarrow c_{12}$ the result differs from 
the fact that we have $3N_{f1}$ branes on
$[c_{11} = c_{12}=c_{1}, a_{2}]$ instead of $2 N_{f1}$. 
We have the following configuration: $3N_{f2}$ D 6-branes are wrapped
on $[c_{2}, a_{11}=a_{12}]$, $(N_{c1} + N_{f2})$ D 6-branes are wrapped on 
$[a_{11}=a_{12}, b]$, $(N_{f1} + N_{c2})$ D 6-branes are wrapped on 
$[b, c_{11} = c_{12}]$
and $3N_{f1}$ D 6-branes are wrapped between $[c_{11}=c_{12}, a_{2}]$. 
We want to move 
to another point of moduli space of the CY threefold by changing the positions 
of $a_{11}=a_{12}=a_{1}$ and $c_{11} = c_{12} = c_{1}$. We push $c_{1}$
and $b$ on the imaginary direction and move them along the negative real 
direction to the right of $a_{1}$ to obtain the magnetic theory. The
final configuration is $(c_{2}, c_{1}, b, a_{1}, a_{2})$ with 
$3N_{f2}$ D 6-branes wrapped on $[c_{2}, c_{1}]$, $(3N_{f2}+2N_{f1}-N_{c2})$
D 6-branes wrapped on $[c_{1}, b]$, $(3N_{f1} + 2N_{f2}- N_{c1})$ D 6-branes
wrapped on $[b, a_{1}]$ and $3N_{f1}$ D 6-branes wrapped on $[a_{1}, a_{2}]$.
This gives a field theory with the gauge group 
$SU(3N_{f2}+2N_{f1}-N_{c2})\times SU(3N_{f1}+2N_{f2}-N_{c1})$ with 
$N_{f1}(N_{f2})$
flavors in the fundamental of the second(first) gauge group.

The generalisation to arbitrary $k$ becomes now obvious. Instead of 
$a_{1}(c_1)$ we 
take $k$ singular points $a_{11}, a_{12}, \cdots, a_{1k}(
c_{11}, c_{12}, \cdots, c_{1k})$. 
The points $a_{2}, b,$ and $ c_{2}$ will remain
in a single copy. Now wrap $N_{i}$ on $[a_{1i}, b]$ for each $i$ 
between indices 1 and $k$ and
$M_{j}$ on $[b, c_{1j}]$ for each $j$ between indices 
1 and $k$ with the condition
$\sum_{i=1}^{k} N_{i} = N_{c1}$ and $\sum_{j=1}^{k} M_{j} = N_{c2}$. Also
we wrap $N_{f1}$ on $[b, a_{2}]$ and $N_{f2}$ on $[c_{2}, b]$. When we 
displace $a_{11}, \cdots, a_{1k}$ to the right of $c_{2}$, 
for every transition,
$N_{f2}$ supplementary D 6-branes will appear. 
Eventually we have $(k+1)N_{f2}$ D 6-branes wrapped on
$[c_{2}, a_{11}]$, etc. 
The same thing happens when we move $c_{11}, \cdots, c_{1k}$
to the left of $a_{2}$ and we finish with $(k+1)N_{f1}$ branes wrapped on 
$[c_{11}, a_{2}]$. Now we identify $a_{11} = \cdots = a_{1k}= a_{1}$ and 
$c_{11}= \cdots = c_{1k}= c_{1}$. 
The configuration is similar with the one of the
previous subsection, the only difference being the fact that we have 
$(k+1) N_{f1}$ D 6-branes on $[c_{2}, a_{1}]$ and 
$ (k+1) N_{f2}$ D 6-branes on $[c_{1}, a_{2}]$ instead of $2 N_{f1}$ and 
$2 N_{f2}$ respectively. We change again the positions of $b$ and $c_{1}$ from 
the right to the left of $a_{1}$ to obtain the magnetic description. 
In terms of field theory this has a gauge group 
$SU((k+1)N_{f2} + kN_{f1}- N_{c2})\times SU((k+1)N_{f1} + kN_{f2}-N_{c1})$
with $N_{f2}(N_{f1})$ flavors in the fundamental of the first(second) 
gauge group. 
The result is the same as the one obtained in \cite{bh} by changing the brane
configuration.

{\bf 2.3) $SU(N_{c1})\times SU(N_{c2})$ with adjoint matter}

Now we introduce adjoint matter. In the brane configuration picture, this 
corresponds to $k$ NS 5-branes with the same orientation as A connected by
$N_{c1}$ D 4-branes with $k$ NS 5-branes with the same orientation as B
which are connected to $k$ NS 5-branes with the same orientation
as C by $N_{c2}$ D 4-branes. The adjoint fields $X_{1}$ and $X_{2}$ 
correspond to the motion
of the $k$ NS 5-branes in $(x^{4}, x^{5}, x^{8}, x^{9})$ directions. In the 
geometrical picture, after a T-duality, this means that we have 
$k$ NS 5-branes at $a_{1}$,
$k$ NS 5-branes at $b$ and $k$ NS 5-branes at $c_{1}$. Using the ideas of 
\cite{ca1,ca2}, we apply the same methods as in section 2.2.
Again we take instead of $k$ NS 5-branes at one point, $k$ points with
one NS 5-brane at each point. This means that instead of single singular
points $a_{1}, b, c_{1}$ we take $k$ copies of them( i.e. we take
as singular points $a_{11}, \cdots, a_{1k}, b_{1}, \cdots, 
b_{k}, c_{11}, \cdots, c_{1k}$
with one NS 5-brane at each point). 
The electric theory has the gauge product $SU(N_{c1})\times SU(N_{c2})$,
with $N_{f1}(N_{f2})$ flavors in the fundamental representation of 
the first(second) gauge group.
Besides we have flavors in the adjoint representations of each component
of the product gauge group. Denote $X_{1}(X_2)$ the adjoint of the 
first(second) gauge group. Then the
superpotential is:
$
W = \mbox{Tr} X_{1}^{k+1} + \mbox{Tr} X_{2}^{k+1} + 
\mbox{Tr} X_{1}YX + \mbox{Tr} X_{2}YX + 
\rho_{1} \mbox{Tr} X_{1} + \rho_{2} \mbox{Tr} X_{2}
$
where $\rho_{i}$ are Lagrange multipliers necessary in order to
impose the tracelessness condition for $X_{1}$ and $X_{2}$. 
The power $k$ can be any positive integer. If we take $k$ to be one, 
the problem
reduces to the one treated in the section 2.1. 
So we will take $k$ greater than 2.

Let us take $k=2$. the configuration looks like 
this: $(a_{11}, a_{12}, c_{2}, b_{1}, b_{2}, a_{2}, c_{11}, c_{12})$.
Note that $a_{2}$ and $ c_{2}$ are always in a single copy. We now wrap
$N_{1}$ D 6-branes on $[a_{11}, b_{1}]$, $N_{2}$ D 6-branes 
on $[a_{12}, b_{2}]$,
$N_{f2}$ on $[c_{2}, b_{1}]$, $N_{f2}$ on $[c_{2}, b_{2}]$, $N_{3}$
on $[b_{1}, c_{11}]$, $N_{4}$ on $[b_{2}, c_{12}]$, $N_{f1}$ on 
$[b_{1}, a_{2}]$, and $N_{f1}$ on $[b_{2}, a_{2}]$, with the condition that
$N_{1} + N_{2} = N_{c1}, N_{3} + N_{4} = N_{c2}$. 
If we take the limit $b_{1}\rightarrow b_{2}$, 
we have now $2 N_{f2}$ D 6-branes 
wrapped on $[c_{2}, b_{1} = b_{2}]$ and $2N_{f1}$ D 6-branes on 
$[b_{1} = b_{2}, c_{1}]$. Again the configuration looks very much alike the
previous ones.

We now move $a_{11}, a_{12}$ to the right of $c_{2}$ and 
$c_{11}, c_{12}$ to the left of $a_{2}$. In this case, we apply again 
the observation 
about the supplementary branes. In the present case the transition
looks as in section 2.2 but we start with $2N_{f2}$ D 6-branes wrapped on
$[c_{2}, b_{1} = b_{2}]$ and $2N_{f1}$ D 6-branes wrapped on
$[b_{1} = b_{2}, a_{2}]$. We obtain $4 N_{f2}$ D 6-branes wrapped
on $[c_{2}, a_{11}]$, $(3N_{f2} + N_{1})$ wrapped on $[a_{11}, a_{12}]$, 
$(2N_{f2} + N_{c1})$ wrapped on  $[a_{12}, b]$, $(2 N_{f1} + N_{c2})$ 
wrapped on
$[b, c_{11}]$, $(3 N_{f1} + N_{4})$ wrapped on $[c_{11}, c_{12}]$ and
$4 N_{f2}$ wrapped on $[c_{22}, a_{2}]$. 

We now move to other point in the moduli of the CY threefold and end up with
the magnetic theory. Taking the limit $a_{12}\rightarrow a_{11}$ and
$c_{12}\rightarrow c_{11}$, we have the following starting configuration
$(c_{2}, a_{11} = a_{12}=a_{1}, b, c_{11} = c_{12}=c_{1}, a_{2})$, with
$4N_{f2}$ D 6-branes wrapped on $[c_{2}, a_{1}]$, $(2N_{f2} + N_{c1})$ wrapped
on $[a_{1}, b]$, $(2N_{f1} + N_{c2})$ wrapped on $[b, c_{1}]$ and 
$4N_{f1}$ wrapped on $[c_{1}, a_{2}]$. To get the magnetic description
we move again $b$ and $c_{1}$ from the right to the left of $a_{1}$. The
procedure is the same as before and we obtain the magnetic
theory with the gauge group $SU(4N_{f2} + 2N_{f1} - N_{c2})\times
SU(4N_{f1} + 2N_{f2} - N_{c1})$. 
Our result again agrees with the result of \cite{Brodie,bh}
for $k=2$.

To generalise to the case of arbitrary $k$, we have to take $k$ copies of all
the singular points $a_{1}, b, c_{1}$. Compared to section 2.2, the only
difference is the appearance of $k$ copies of $b$. When all these copies 
coincide, we will have $kN_{f2}$ D 6-branes wrapped on $[c_{2}, b]$ and
$kN_{f1}$ wrapped on $[b, a_{1}]$. The electric theory has the following
configuration of singular points $(c_{2}, a_{11} = \cdots = a_{1k} = a_{1}, 
b_{1} = \cdots = b_{k} = b, c_{11} = \cdots = c_{1k} = c_{1}, a_{2})$, 
with $2kN_{f2}$ D 6-branes wrapped 
on $[c_{2}, a_{1}]$, $(kN_{f2} + N_{c1})$ D 6-branes wrapped
on $[a_{1}, b]$, $(kN_{f1} + N_{c2})$ D 6-branes wrapped on $[b, c_{1}]$
and $2kN_{f1}$ D 6-branes wrapped on $[c_{1}, a_{2}]$.   
The magnetic theory is obtained by changing the position of $b$ 
and $c_{1}$ from
right to left with respect to $a_{1}$. The procedure is the same as the one
performed in section 2.1 and we obtain the magnetic theory with the
gauge group $SU(2kN_{f2} + kN_{f1} - N_{c2})\times 
SU(2kN_{f1} + k N_{f2} - N_{c1})$, result which again agrees with the one 
obtained in field theory and in brane configuration for 
any value of $k$.

\section{The product of more than two SU groups}

To complete the analogy with the results obtained by field theory methods and
by brane configurations, we consider now the case of a product of more than
two SU groups. One of the most important features of 
our construction in this case
is that we were able to give an explanation in the geometric picture for
the semi-infinite D 4-branes that appear in the brane configuration picture.
In \cite{bh}, these semi-infinite D 4-branes were neccesary in order to match
the field theory calculation for the dual gauge group. In our case, we make
an assumption based on 
the dualities in geometrical picture. That is, for $SU(N_{c})$
case, the flavor group in the dual theory is determined by the D 6-branes 
which are wrapped on $[a_{1}, a_{2}]$. For $SU(N_{c1})\times SU(N_{c2})$
case, the flavor group in the dual picture is determined by the 
D 6-branes which
are wrapped on $[a_{1}, a_{2}]$ and $[c_{1}, c_{2}]$. We thus observe that
the flavor group in the dual theory is always determined by D 6-branes
wrapped on pair of singular points. By pair of singular points we understand 
here points where, after a T-duality, the NS 5-branes are parallel.
This is a main observation and we will use it to obtain the gauge group in the
magnetic theory. We will also show that this condition is completely
equivalent to the requirement of introducing semi-infinite D 4-branes!

{\bf 3.1) $SU(N_{c1})\times SU(N_{c2})\times SU(N_{c3})$ }

Consider the theory with gauge group $SU(N_{c1})\times SU(N_{c2})\times 
SU(N_{c3})$, with $N_{f1}(N_{f2})[N_{f3}]$ 
flavors in the fundamental representation of the
first(second)[third] SU group. 
Besides, we have a field X in the 
$\bf{ (N_{c1}, \overline{N}_{c2})}$
representation and its conjugate $\widetilde{X}$ and a field Y in the
$\bf {(N_{c2}, \overline{N}_{c3})}$ 
representation and its conjugate $\widetilde{Y}$.
To truncate the chiral ring, we add the superpotential:
$
W= \frac{1}{2} \mbox{Tr} (X\widetilde{X})^{2} + 
\mbox{Tr} X\widetilde{X}Y\widetilde{Y}-
 \frac{1}{2} \mbox{Tr} (Y\widetilde{Y})^{2}.
$
Now we want to see how this looks like in the geometric picture. Consider the
singular points to be, from left to right $(a_{1}, c_{2}, b, d_{2}, a_{2},
c_{1}, d_{1})$. We wrap $N_{c1}$ D 6-branes on $[a_{1}, b]$, $N_{c2}$
D 6-branes on $[b, c_{1}]$, $N_{c3}$ D 6-branes on $[c_{1}, d_{1}]$,
$N_{f1}$ D 6-branes on $[b, a_{2}]$, $N_{f2}$ D 6-branes on $[c_{2}, b]$ and
$N_{f3}$ D 6-branes on $[d_{2}, c_1]$ which is the electric theory.

We do want to go to the magnetic theory by moving 
to another point in the moduli
of the CY threefold. First move $c_{2}$ and $d_{2}$ to the left
of $a_{1}$ in order to be able to send them to infinity and to obtain
a global flavor group.  When moving $c_{2}$ to the
left of $a_{1}$, there will be $N_{f2}$ {\it additional} D 6-branes wrapped on
$[c_{2}, a_{1}]$. So there will be $2N_{f2}$ D 6-branes wrapped on
$[c_{2}, a_{1}]$. Now move the point $d_{2}$. 
When $d_{2}$ passes $b$, there are
$N_{f3}$ {\it additional} 
D 6-branes wrapped on $[d_{2}, b]$. When $d_{2}$ passes
$a_{1}$ there are another $N_{f3}$ additional D 6-branes wrapped on
$[d_{2}, a_{1}]$ which one has to add to the previous additional $N_{f3}$.
Insert now $d_{2}$ between $c_{2}$ and $a_{1}$.
Then we have $2N_{f2}$ D 6-branes wrapped on $[c_{2}, d_{2}]$, 
$(2N_{f2} + 3N_{f3})$ wrapped on $[d_{2}, a_{1}]$ and $(2N_{f3} + N_{c1} +
N_{f2})$ wrapped on $[a_{1}, b]$. 
We now move $a_{2}$ to the right of $c_{1}$ and $d_{1}$. When $a_{2}$ passes
$c_{1}$, there are $N_{f1}$ additional D 6-branes wrapped on $[c_{1},a_{2}]$.
When $a_{2}$ passes $d_{1}$ there another additional $N_{f1}$ D 6-branes
giving $2N_{f1}$ additional D 6-branes wrapped on $[d_{1}, a_{2}]$. 
So, when we move $a_{2}$ to the right of $c_{1}$ and $d_{1}$, we have
$(N_{f1} + N_{c2} + N_{f3})$ D 6-branes wrapped on $[b, c_{1}]$,
$(2N_{f1} + N_{c3})$ wrapped on $[c_{1}, d_{1}]$ and $3N_{f1}$ wrapped on
$[d_{1}, a_{2}]$. 

We now move to the magnetic theory. To do that, we move $b$, 
$c_{1}$ and $d_{1}$
to the left of $a_{1}$, obtaining the configuration 
$(c_{2}, d_{2}, d_{1}, c_{1}, b, a_{1}, a_{2})$ on the real axis. 
We first move the point $d_{1}$. 
When we push $d_{1}$ along the imaginary direction,
there are $(2N_{f1} + N_{c3})$ D 6-branes 
wrapped directly on $[c_{1}, a_{2}]$ and $(N_{f1} - N_{c3})$ on 
$[a_{2}, d_{1}]$.
By moving $d_{1}$ along the negative real axis and inserting it between 
$d_{2}$ and $a_{1}$, we obtain $(3N_{f3} + 2N_{f2} + N_{f1} - N_{c3})$
D 6-branes wrapped on $[d_{1}, a_{1}]$, $(2N_{f3} + N_{f1} + N_{f2} + N_{c1}-
N_{c3})$ on $[a_{1}, b]$, $(2N_{f1} + N_{f3} + N_{c2} - N_{c3})$ on
$[b, c_{1}]$ and $3N_{f1}$ on $[c_{1}, a_{2}]$. 
 When we push $c_{1}$ along the imaginary direction,
there are $(N_{f1} - N_{c2} - N_{f3} + N_{c3})$ D 6-branes wrapped on
$[c_{1}, b]$ and $(2N_{f1} + N_{f3} + N_{c2} - N_{c3})$ D 6-branes wrapped
on $[b, a_{1}]$. The point 
$c_{1}$ is inserted between $d_{1}$ and $a_{1}$ so there
will be $(2N_{f3} + 2N_{f2} + 2N_{f1} - N_{c2})$ D 6-branes wrapped on
$[c_{1}, a_{1}]$. The last step is to move $b$ between $c_{1}$ and $a_{1}$.
By making this we  obtain that there are $(3N_{f1} + N_{f2} + N_{f3} - N_{c1})$
D 6-branes wrapped on $[b, a_{1}]$ and $3N_{f1}$ D 6-branes wrapped on 
$[a_{1}, a_{2}]$. The magnetic picture that we have obtained is identical with
the one obtained in \cite{bh}. But it does not match exactly the one 
obtained by field theory methods. 

In \cite{bh}, they have obtained the
field  theory result by introducing semi-infinite D 4-branes. To apply their 
method, we first make a T-duality transformation. We have NS 5-branes at each 
of the singular points. The NS 5-branes at $a_{2}, c_{2}, d_{2}$ play the role
of the $N_{f1}, N_{f2}$ and $N_{f3}$ D 6-branes in brane configuration
picture. Consider that we did not send $a_{2}, c_{2}$ and $d_{2}$ to infinity.
Insert $2N_{f2}$ D 4-branes between the NS 5-branes 
sitting at $c_{1}$ and $c_{2}$.
The conservation of the linking number for the NS 5-branes sitting at $c_{2}$
requires $2N_{f2}$ semi-infinite D 4-brane at its left. The conservation of
the linking number for the NS 5-brane sitting at $c_{1}$ requires $2N_{f2}$
D 4-branes at its right. They will combine with the $(2N_{f2} + 2N_{f3}
+ 2N_{f1} - N_{c2})$ D 4-brane existent between the NS 5-branes sitting at 
$c_{1}$ and $b$ to give $(4N_{f2} + 2N_{f3} + 2N_{f1} - N_{c2})$ D 4-branes 
which is the
right result for the middle gauge group obtained by field
theory methods. Now we have to add $2N_{f2}$
D 4-branes to the right of the NS 5-brane sitting at the point $b$. 
$N_{f2}$ of them combine
with the rest of D 4-branes between the NS 5-branes sitting at the point $b$ 
and $a_{1}$ to give $(3N_{f1} + 2N_{f2} + N_{f3} - N_{c1})$ D 4-brane which
is the right result for the last part of the product gauge group. To
conserve the linking number, we need to add $N_{f2}$ semi-infinite 
D 4-branes
to the right of the NS 5-branes sitting at $b$ and $a_{1}$. 

Let us see what is the correspondence in the original geometric picture.
Before any insertion, we have $3N_{f3}$ D 6-branes wrapped on $[d_{2}, d_{1}]$
and $2N_{f2}$ D 6-branes wrapped on $[c_{2}, d_{1}]$. As we discussed at
the beginning of this section, we put the condition that in the
dual picture, the flavor group is given by D 6-branes wrapped only on
the pair singularity points. 
 We have cycles of pairs $[c_{1}, c_{2}]$, $[a_{1}, a_{2}]$ and
$[d_{1}, d_{2}]$. The point $b$ 
does not have a pair, so we might have D 6-branes
wrapped for example on $[b, a_{2}]$.

After the transition, we have D 6-branes wrapped on $[a_{1}, a_{2}]$ and
$[d_{1}, d_{2}]$ but we also have  $2N_{f2}$ D 6-branes wrapped
on $[c_{2}, d_{1}]$ which do not satisfy our condition. We want to
extend these to D 6-branes wrapped on $[c_{2}, c_{1}]$. To completely 
understand what is the process, we need to go back and forth
between the original geometrical picture and its T-dual. The insertion of
$2N_{f2}$ D 4-branes between the NS 5-branes sitting at $c_{1}$ and 
$c_{2}$ means here to connect the D 6-branes wrapped on $[c_{2}, d_{1}]$
with $2N_{f2}$ wrapped on $[d_{1}, c_{1}]$. The number of D 6-branes
wrapped on $[d_{1}, c_{1}]$ remains the same (and gives 
the right gauge product)
but we have now $2N_{f2}$ D 6-branes wrapped on $[c_{2}, c_{1}]$.
The supplementary $2N_{f2}$ D 4-branes 
appearing in the dual picture between the
NS 5-branes sitting at $b$ and $a_{1}$ just represent $2N_{f2}$ supplementary
D 6-branes wrapped on $[b, a_{1}]$ and they add to the $(2N_{f3} + 2N_{f2} + 
2N_{f1} - N_{c2})$ D 6-branes wrapped after the transition on the same cycle.
So we have $(4N_{f2} + 2N_{f3} + 2N_{f1} - N_{c2})$ D 6-branes wrapped on 
$[c_{1}, b]$ which gives the correct result for the middle gauge group.
Also the final gauge group is correct when we take the dual back to the
original geometrical picture. We need $N_{f2}$ supplementary D 6-branes
wrapped on $[b, a_{2}]$. This implies that we have
$N_{f2}$ supplementary D 6-branes wrapped on $[b, a_{1}]$  leading to 
$(3N_{f1} + 2N_{f2} + N_{f3} - N_{c1})$ D 6-branes wrapped on
$[b, a_{1}]$ and this gives the right result for the last gauge group.

So, after transition, by wrapping $N_{f2}$ D 6-branes on $[b, a_{2}]$ and
by wrapping $2N_{f2}$ D 6-branes on $[c_{2}, c_{1}]$ instead of
$[c_{2}, d_{1}]$ we obtain the correct magnetic theory, which has 
gauge group $SU(3N_{f3} + 2N_{f2} + N_{f1} - N_{c3})\times
SU(4N_{f2} + 2N_{f3} + 2N_{f1} - N_{c2})\times SU(3N_{f1} + 2N_{f2} + N_{f3}
- N_{c1})$. The additional D 6-branes came from the condition stated
in the first paragraph of this section. 
This is the result obtained in \cite{bh} and also obtained by field
theory methods. 

{\bf 3.2) $SU(N_{c1})\times SU(N_{c2}) \times SU(N_{c3})$ 
with adjoint matter}

Now we introduce adjoint matter. The electric theory has the gauge group
$SU(N_{c1})\times SU(N_{c2})\times SU(N_{c3})$ with $N_{f1}(N_{f2})[N_{f3}]$ 
flavors in the
fundamental representation of the first(second)[third] 
gauge group. 
Besides we have flavors in
the adjoint representations of each of 
the components of the gauge group. Denote
by $X_{1}(X_2)[X_3]$ the adjoint of the first(second)[third] 
gauge group.
The superpotential which truncates the chiral ring is:
$
W  = 
\mbox{Tr} X_{1}^{k+1} + \mbox{Tr} X_{2}^{k+1} + \mbox{Tr} X_{3}^{k+1} + 
\mbox{Tr} X_{1}X \widetilde{X}
+ \mbox{Tr} X_{2}Y \widetilde{Y} + \mbox{Tr} X_{3}Y\widetilde{Y} 
  + \mbox{Tr} X_{2}X\widetilde{X}
+\rho_{1} \mbox{Tr} X_{1} + \rho_{2} \mbox{Tr} X_{2} + 
\rho_{3} \mbox{Tr} X_{3}
$
where again $\rho_{i}$ are Lagrange multipliers to enforce the tracelessness 
conditions for $X_{i}$. The power $k$ can be any positive integer.

In the brane configuration picture, 
 this corresponds 
to $k$ NS 5-branes with the same orientation as A connected by $N_{c1}$
D 4-branes with $k$ NS 5-branes with the same orientation as B. These are
connected by $N_{c2}$ D 4-branes with $k$ NS 5-branes with the same orientation
as C and the last ones are connected by $N_{c3}$ D 4-branes 
with $k$ NS 5-branes
with the same orientation as D. In the geometric picture,
this means that we take $k$ copies of all the
singular points, except for $a_{2}, c_{2}$ and $d_{2}$. After moving 
$c_{2}, d_{2}$ to the left of all copies of $a_{1}$ and $c_{1}, 
d_{1}$ to the left of all copies of 
$a_{2}$, we obtain $2kN_{f2}$ D 6-branes wrapped on $[c_{2}, d_{2}]$ , 
$(2kN_{f2} + 3kN_{f3})$ D 6-branes wrapped on $[d_{2}, a_{1}]$,
$(2kN_{f3} + kN_{f2} + N_{c1})$ D 6-branes wrapped on $[a_{1}, b]$,
$(kN_{f1} + kN_{f3} + N_{c2})$ D 6-branes wrapped on $[b, c_{1}]$,
$(2kN_{f1} + N_{c3})$ D 6-branes wrapped on $[c_{1}, d_{1}]$ and $3kN_{f1}$
D 6-branes wrapped on $[d_{1}, a_{2}]$. 

We now move to the magnetic theory. The dual configuration is given by 
$k$ copies of the 
one discussed in the previous subsection. This can be seen by explicitly
exchanging the positions of $b, c_{1}, d_{1}$ from right to left with 
respect to $a_{1}$. Again we obtain  $2kN_{f2}$
wrapped on $[c_{2}, d_{1}]$ and we want to extend them to
 D 6-branes on $[c_{2}, c_{1}]$
instead of $[c_{2}, d_{1}]$. This is required by the condition that the
flavor group in the dual picture is given by D 6-branes wrapped on
the pair singularity points. This is 
again equivalent to introducing semi-infinite D 4-branes in brane configuration
picture, in order to match the field theory result. 
The magnetic theory has a gauge group $SU(3kN_{f3} + 2kN_{f2} + kN_{f1} - 
N_{c3})
\times SU(4kN_{f2} + 2kN_{f1} + 2kN_{f3} - N_{c2})\times SU(3kN_{f1} + 
2kN_{f2} + kN_{f3} - N_{c1})$ which coincides again with the one obtained 
in brane configuration picture.

{\bf 3.3) Generalisation to arbitrary number of product gauge groups}

Consider now the electric theory with the gauge group $SU(N_{c1})\times
SU(N_{c2})\times SU(N_{c3}) \times \cdots \times SU(N_{cn})$. Each gauge group
has fundamental fields $N_{f1}, N_{f2}, \cdots, N_{fn}$. Besides we have
the fields $Y_{1}$ in the representation $\bf {(N_{c1}, 
\overline{N}_{c2}, 1, \cdots, 1)}$,
$Y_{2}$ in the representation $\bf {(1, N_{c2}, \overline{N}_{c3}, 1, \cdots, 
1)}, \cdots,
Y_{n-1}$ in the representation $\bf {(1, 1, \cdots, 
N_{cn-1},\overline{N}_{cn})}$ and their
conjugate fields $\widetilde{Y}_{1}, \cdots, \widetilde{Y}_{n-1}$. 
We also need
a superpotential to truncate the chiral ring.

We want to see how this looks in the geometrical picture. Take for
simplicity $n=4$. Consider the singular
points to be, from left to right $(a_{1}, c_{2}, b, d_{2}, a_{2}, c_{1}, e_{2},
d_{1}, e_{1})$. We wrap $N_{c1}$ D 6-branes on $[a_{1}, b]$, $N_{c2}$ 
D 6-branes
on $[b, c_{1}]$, $N_{c3}$ D 6-branes on $[c_{1}, d_{1}]$, $N_{c4}$ 
D 6-branes
on $[d_{1}, e_{1}]$, $N_{f1}$ D 6-branes on $[b, a_{2}]$, 
$N_{f2}$ D 6-branes on
$[b, c_{2}]$, $N_{f3}$ D 6-branes on $[c_{1}, d_{2}]$ and $N_{f4}$ D 6-branes
on $[d_{1}, e_{2}]$ which is the electric theory. 
Before moving to the magnetic
theory, we move $c_{2}, d_{2}$ and $e_{2}$ to the left of $a_{1}$ and
$c_{1}, d_{1}$ and $e_{1}$ to the left of $a_{2}$. Some additional 
D 6-branes
will be wrapped after each transition. We obtain, from left to right,
$2N_{f2}$ D 6-branes wrapped on $[c_{2}, d_{2}]$, $(2N_{f2} + 3N_{f3})$ 
D 6-branes
wrapped on $[d_{2}, e_{2}]$, $(2N_{f2} + 3N_{f3} + 4N_{f4})$ D 6-branes
wrapped on $[e_{2}, a_{1}]$, $(3N_{f1} + N_{c3})$ D 6-branes wrapped on 
$[d_{1}, e_{1}]$ and $4N_{f1}$ D 6-branes wrapped on $[e_{1}, a_{2}]$.

We now move to the magnetic theory. We push first $e_{1}$ between $e_{2}$
and $a_{1}$, so there are $\widetilde{N}_{c1}( = 4N_{f4} + 3N_{f3} + 2N_{f2} + 
N_{f1} - N_{c4})$ D 6-branes wrapped on $[e_{1}, a_{1}]$. This is the
right result for the first dual gauge group in the product. We subsequently
move $d_{1}$, $c_{1}$ and $b$ 
to the right of $a_{1}$. In the final picture, the
gauge group that we obtain is not the same as the one obtained by field theory
methods. We can match it by introducing semi-infinite D 4-branes in brane
configuration picture or we can match it by imposing that the D 6-branes
which do not contribute to the gauge products are
wrapped on 3-cycles between the pairs of points, $[a_{1}, a_{2}]$,
$[c_{1}, c_{2}]$, $[d_{1}, d_{2}]$ and $[e_{1}, e_{2}]$. 
In this case, we have obtained the
magnetic theory with the gauge group $SU(4N_{f4} + 3N_{f3} + 2N_{f2} + N_{f1}-
N_{c4})\times SU(3N_{f4} + 6N_{f3} + 4N_{f2} + 2N_{f1} - N_{c3}) \times
SU(2N_{f4} + 4N_{f3} + 6N_{f2} + 3N_{f1} - N_{c2}) \times SU(N_{f4} + 2N_{f3}
+ 3N_{f2} + 4N_{f1} - N_{c1})$, with $N_{f4}$ flavors in the fundamental
of the first gauge group, etc. This again agrees with the results of \cite{bh}
for $n=4$.

In order to obtain a generalisation to any value of $n$, one has to take 
$(2n-1)$ singular points and to move them from right to left with respect to
a reference point. We always have to be able to extend the flavor cycles to 
infinity. So we always push $a_{2}$ to the far right and $c_{2}, d_{2},
\cdots$
to the far left. In order to obtain the result of field theory, we 
have to impose that the D 6-branes which do not contribute to the
gauge group are to be wrapped on 3-cycles between pairs of points, like
$[a_{1}, a_{2}]$, etc. This is equivalent with introducing the semi-infinite
D 4-branes in brane configuration method. 
Our result again agrees with the one
of \cite{bh}.
It is also possible to
add adjoint matter fields, this imposing the multiplicity of each 
singular point, so the theory with adjoint is just made out of several
copies of the theory without adjoint matter and the procedure is the same as
before.

\section{Conclusion}

We have seen that a number of $N=1$ supersymmetric field theory dualities
were obtained in terms of geometric realization of wrapping D 6-branes
around 3-cycles of CY threefold in type IIA string theory.
The condition that the flavor gauge group in the dual theory is
determined by D 6-branes wrapping around only {\it pairs} of singular points
proved to be crucial for our construction. Our construction 
also gives rise to extra D 6-branes
wrapping around the cycles.
It would be interesting to study this transition further.  

{\bf Notes added}: Recent work \cite{BDG} on the creation of D-branes
might be related to our discussion for the supplementary D 6-branes.


CA thanks Jaemo Park for the correspondence and
RT would like to thank Orlando Alvarez for advices.

\end{document}